# Influence of Na-doping and sintering temperature on increasing $Bi_2Sr_2CaCu_2O_8$ superconducting phase content in powder form materials

S. Rahier [a,*], S. Stassen [a], R. Cloots [a] and M. Ausloos [b]

[a] *Chemistry Institute B6, SUPRATECS, University of Liège, Sart-Tilman, B-4000 Liège, Belgium*
[b] *Physics Institute B5, SUPRATECS, University of Liège, Sart-Tilman, B-4000 Liège, Belgium*

**Abstract**

This paper proposes a systematic way to explore the effects of Na on Bi-2212 phase formation. The influence of sodium metal is correlated with the 2212 fraction formed in the sample. The effect on the sintering temperature and the importance of the substitution site is studied. Samples are characterized by X-ray diffraction analysis (XRD). Results clearly show that the 2212 phase content is enhanced and sintering temperatures can be lowered with regard to the undoped samples.

## 1. Introduction

Bi-based high-temperature superconducting phases, represented by $Bi_2Sr_2Ca_{n-1}Cu_nO_{2n+4}$ (n = 1, 2 or 3) and written 2201, 2212 and 2223 have an increasing value of their superconducting critical temperature. The 2212 phase is the most familiar one because it can be synthesized as a single-phase material.

As for YBCO materials [1, 2], it is of great interest to study the effects of Na ion substitution in 2212 materials since beside providing some information on the chemical and physical properties of the system, it might lead to an improvement of their technological properties, such as the superconducting critical temperature and the current carrying capacity.

Na-doped Bi-2212 compounds have already been investigated by several groups [3-6]. The work of Dou *et al*. [3] is devoted to the mechanism of $T_c$ enhancement by sodium doping in $Bi_{2.2}Sr_{1.8}Ca_{1.05}Cu_{2.15-y}Na_yO_{8+\delta}$ samples. A superconducting critical transition temperature of 94 K and a $T_{c\ off}$ (the temperature at which R(T) = 0) of 90 K has been achieved. Moreover Dou *et al.* observed that the *a*- and *c*-axes increased while the *b*-axis decreased with increasing the Na content, indirectly indicating that Na might have been introduced into the lattice even though it was not found on which crystallographic site Na was located.

The research of Yu *et al*. [4] described the preparation of $Bi_{2.1}Sr_{1.8}Ca_{1.05}Na_yCu_{2.15-y}O_{8+\delta}$ (y = 0, 0.4, 0.5, 0.6, 0.7, 0.8, 1.0) samples using melt processing followed by quenching. The highest $T_c$ achieved was 94 K. During annealing in flowing oxygen, $T_c$ did not change. But $T_c$ was degraded under high oxygen pressure annealing. NMR analysis revealed that Na might have replaced Bi and Ca atoms.

Chandra Sekhar *et al*. [5], have studied the ultrasonic longitudinal velocity on sodium-doped Bi-2212 single-phase pepared by solid state route using the pulse transmission technique (80 – 300 K). In the temperature range 300 - 230 K, in contrast to normal solids, the velocity of Bi - Na(0) and Bi - Na(3) samples was found to decrease with decreasing temperature (elastic softening) followed by a velocity maximum at around 150 K for all the four samples, signifying the presence of lattice instabilities.

*Corresponding author. Tel. : +32-4-366-3417; fax : + 32-4-366-3413
E-mail address : s.rahier@ulg.ac.be (S. Rahier)

Continuing preliminary studies in [6], the influence of 2212 phase formation to obtain the highest 2212 phase volume fraction is hereby investigated with regards to the Na substitution site and the sintering temperature. In Section 2, experimental procedures are given, i.e. how Na-doped 2212 powders with a $Bi_{2-u}Sr_{2-v}Ca_{1-w}Cu_{2-x}Na_yO_z$ stoichiometry have been prepared. In order to investigate the best substitution location, different mixtures have been realized in which the 2212 phase was substituted on the different possible sites according to the $Bi_{2-u}Sr_{2-v}Ca_{1-w}Cu_{2-x}Na_yO_z$ formula, the u, v, w and x indices indicating the deviations from the classical stoichiometry. The index y is the Na-substitution level. A systematic study of the influence of the sintering temperature and the nature of the substitution site on the formation of the 2212 phase in powder form has been considered. The samples have been characterized by XRD. In Section 3, the results are summarized. The last section is devoted to the concluding remarks.

## 2. Experimental details

The $Bi_{2-u}Sr_{2-v}Ca_{1-w}Cu_{2-x}Na_yO_z$ powder synthesis starts by mixing appropriate amounts of $Bi_2O_3$, $SrCO_3$, $CaCO_3$, $CuCO_3.Cu(OH)_2$ and $NaHCO_3$. The mixture was treated at 620°C during 24 h, for decarbonization, and was then sintered at a temperature $T_1$, during 24 h with a heating rate of 150 °C/h. Six $T_1$ values regularly ranging from 750 to 830 °C were considered (Table 1). The powders were characterized by X-ray diffraction for phase determination and quantification. The Na substitution level i.e. y and one of the (u, v, w, x) are fixed at 0.3, the other three deviation indices were being taken equal to 0.0. For comparison the 4 corresponding cases with (u, v, w, x) = 0.0 still with y = 0.3 were examined. Moreover, a classical Bi-2212 has been synthesized giving 10 samples to analyse (Table 1).

## 3. Results and discussion

Each 2212 to the 2201 phase content has been evaluated by means of a X-ray diffraction peak intensity measurement on the (103) peak taking into account the intensity of the (105) peak as the reference one for representing a theoretical single-phase material (see Table 2).

The relative volume fractions of the low-$T_c$ ($f_{2212}$) and non-superconducting ($f_{2201}$) phases were determined from such peak intensities as follows. Define the fraction of the 2212 phase to be given by $f_{2212}$ = 2212 /[2212+2201], and calculate $f_{2212}$ in three different ways, i.e.

- $f_{2212} = [I_{2212} (105)] / [I_{2212} (105) + I_{2201} (105)]$
- $f_{2212} = [I_{2212} (103) /327] / [ (I_{2212} (103) /327) + (I_{2201} (103) /704)]$
- $f_{2212} = [I_{2212} (105) /1000] / [ (I_{2212} (105) /1000) + (I_{2201} (103) /704) ]$

from each peak intensity as given in Table 2. The average $f_{2212}$ values obtained from the three equations were so computed for each sample sintered at each $T_1$ (Table 3). This allows us to discuss the influence of the sintering temperature for the various cases and examine different possibilities of experimental interest.

The numerical data from Table 3 can be useful displayed in a graphical way (Figs. 1-2) for further discussions. Figures 1 and 2 show the 2212 phase content depending on the sintering temperature for undoped and Na-doped 2212 samples respectively in a three dimensional plot. Each location is equivalent in every figure and concerns one particular substitution site and one sintering temperature. The 2212 phase content scale range in Fig. 1 for the undoped case was intentionally fixed at 90 % in order to facilitate the comparison with the doped samples. The display indicates a common feature to the entire set of results, i.e. a decreasing of the 2212 phase formation temperature with sodium metal doping. The maximum 2212 phase content is of the order of 32 % for undoped materials treated at 830 °C whereas the equivalent maximum value reaches as high as 87 % for the same sintering temperature and time for the #8 Na-doped sample.

Thereafter, the influence of the substitution site can be put into evidence. Small variations in the stoichiometry used to grow the undoped 2212 samples have not drastically modified the quantity of the 2212 phase as confirmed by observing the data in Fig. 2. Nevertheless, one has to note that a perfect 2212 stoichiometry, i. e., sample #1, has led to a higher 2212 phase content than other cases, i.e. when one ion was deficient in the formula. The situation is quite different for doped materials for which the 2212 content results depend on the nature of the substitution sites. When we compare results obtained for Na-doping (Fig. 2) at various substitution sites, there is a small advantage from the phase content point of view when there is addition. The worse results being for the case of Ca or Cu deficiency.

## 4. Conclusions

We have enlightened the influence of the Na substitution and the effect of the possible substitution site on the formation temperature of the 2212 phase prepared in powder form. The use of Na ions in the starting mixture seems to decrease the 2212 phase formation temperature and to increase the 2212 phase content. To conclude from observations, we can state that the 2212 phase content, at a given temperature and with Na substitution, is higher when the alkali metal is added than when it substitutes one ion site in the initial stoichiometry, and this stands whatever the substitution site considered.

## 5. References

[1] M. Pekala, H. Bougrine, J. Azoulay and M. Ausloos, Supercond. Sci. Technol. 8, 660-666, **1995**,

[2] R. Cloots, A. Rulmont, M. Pekala, J.Y. Laval, H. Bougrine and M. Ausloos, Z. Phys. B 96, **1995**, 319-324.

[3] S.X. Dou, W.M. Wu, H.K. Liu, C.C. Sorrell, Physica C 185-189, **1991**, 811.

[4] Y. Yu, X. Jin, D.X. Cal, X.X. Yao, C. Hu, K.Y. Ding, D. Feng, Phys. Status Solidi A 146, **1994**, K33.

[5] M. Chandra Sekhar, B. Gopala Krishna, R. Ravinder Reddy, P. Venugopal Reddy and S. V. Suryanarayana, Supercond. Sci. Technol., 9, **1996**, 29-33

[6] S. Stassen, A. Rulmont, M. Ausloos, and R. Cloots, J. Low Temp. Phys. 105, **1996**, 1523 - 1528

**List of captions**

Table 1.
2212 mixtures hereby investigated. The initial chemical composition followed the $Bi_{2-u}Sr_{2-v}Ca_{1-w}Cu_{2-x}Na_yO_z$ formula with the values for the u, v, w, x, and y parameters as given

Table 2.
Peak intensity (%) of particular reflections in the X-ray diffraction patterns for the 2201 and 2212 phases

Table 3.
2212 phase content (in %) of each sample after being treated at $T_1$

Figure 1.
2212 phase content (%) relative to the 2201 phase as a function of the stoichiometry deficiency site (indicated by a minus sign in the 2212 formulae) and the sintering temperature for undoped 2212 samples

Figure 2.
2212 phase content (%) relative to the 2201 phase as a function of the substitution site and the sintering temperature for Na-doped 2212 samples

**List of illustrations**

|  | u=v=w=x=0 | v=w=x=0 u=0.3 (-Bi) | u=w=x=0 v=0.3 (-Sr) | u=v=x=0 w=0.3 (-Ca) | u=v=w=0 x=0.3 (-Cu) |
|---|---|---|---|---|---|
| y=0 | #1 2212 | #2 2212-Bi | #3 2212-Sr | #4 2212-Ca | #5 2212-Cu |
| Na y=0.3 | #6 2212+Na | #7 2212-Bi+Na | #8 2212-Sr+Na | #9 2212-Ca+Na | #10 2212-Cu+Na |

Table 1

| (hkl) | 2201 phase | 2212 phase |
|---|---|---|
| (105) | 100 | 100 |
| (103) | 70.4 | 32.7 |

Table 2

| Sample # | $T_1 =$ 780 °C | $T_1 =$ 790 °C | $T_1 =$ 800 °C | $T_1 =$ 810 °C | $T_1 =$ 820 °C | $T_1 =$ 830 °C |
|---|---|---|---|---|---|---|
| 1 | 12.3 | 15.6 | 18.0 | 24.1 | 27.2 | 32.5 |
| 2 | 7.7 | 10.9 | 20.3 | 17.8 | 23.2 | 25.9 |
| 3 | 10.6 | 16.6 | 17.4 | 17.3 | 24.4 | 22.1 |
| 4 | 7.1 | 12.2 | 16.3 | 17.0 | 21.0 |  |
| 5 | 7.6 | 11.3 | 13.8 | 18.7 | 21.8 | 30.2 |
| 6 | 10.9 | 44.2 | 70.1 | 79.5 | 82.0 | 84.0 |
| 7 | 10.7 | 49.1 | 71.9 | 76.3 | 80.1 | 83.9 |
| 8 | 8.8 | 39.2 | 70.1 | 68.4 | 85.3 | 86.8 |
| 9 | 7.4 | 14.8 | 26.7 | 38.2 | 68.9 | 71.8 |
| 10 | 8.2 | 30.6 | 60.3 | 70.1 | 71.6 | 72.1 |

Table 3

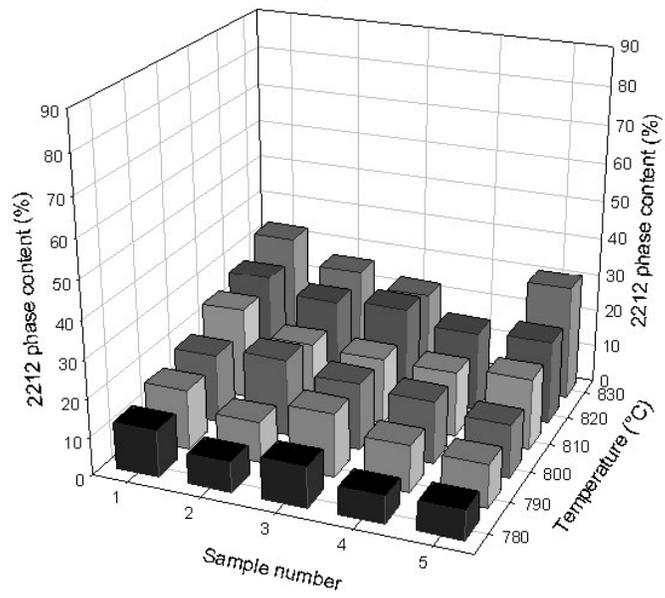

Figure 1

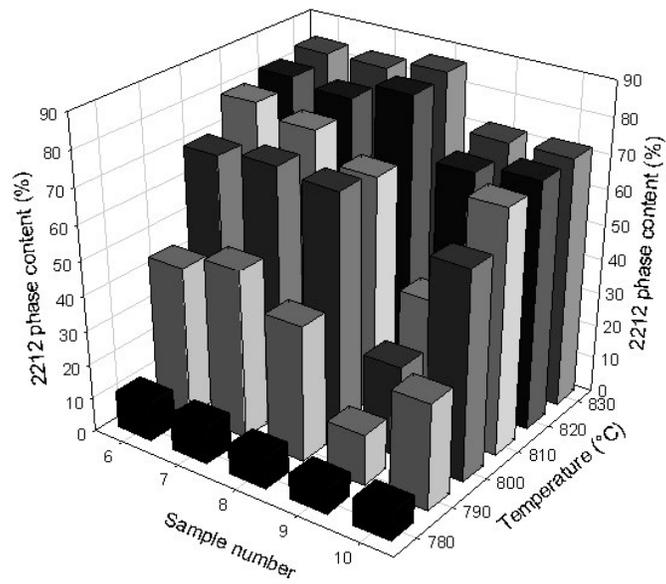

Figure 2